\documentclass{elsart}
\usepackage{ifpdf}
\usepackage{graphicx,natbib,amssymb,lineno}
\ifpdf
\usepackage[%
  pdftitle={Instructions for use of the document class
    elsart},%
  pdfauthor={Simon Pepping},%
  pdfsubject={The preprint document class elsart},%
  pdfkeywords={instructions for use, elsart, document class},%
  pdfstartview=FitH,%
  bookmarks=true,%
  bookmarksopen=true,%
  breaklinks=true,%
  colorlinks=true,%
  linkcolor=blue,anchorcolor=blue,%
  citecolor=blue,filecolor=blue,%
  menucolor=blue,pagecolor=blue,%
  urlcolor=blue]{hyperref}
\else
\usepackage[%
  breaklinks=true,%
  colorlinks=true,%
  linkcolor=blue,anchorcolor=blue,%
  citecolor=blue,filecolor=blue,%
  menucolor=blue,pagecolor=blue,%
  urlcolor=blue]{hyperref}
\fi

\makeatletter
\def\elsartstyle{%
    \def\normalsize{\@setfontsize\normalsize\@xiipt{14.5}}
    \def\small{\@setfontsize\small\@xipt{13.6}}
    \let\footnotesize=\small
    \def\large{\@setfontsize\large\@xivpt{18}}
    \def\Large{\@setfontsize\Large\@xviipt{22}}
    \skip\@mpfootins = 18\p@ \@plus 2\p@
    \normalsize
}
\@ifundefined{square}{}{}
\makeatother

\begin{document}

\begin{frontmatter}

\title{The non-equilibrium work relation. Thermodynamic analysis and microscopic
foundations}
\author{I. Santamar\'{\i}a-Holek \dag, A. P\'{e}rez-Madrid$^{*}$ }
\address{\dag Facultad de Ciencias, Universidad Nacional Aut\'{o}noma de M\'{e}xico.\\
Circuito exterior de Ciudad Universitaria. 04510, D. F., M\'{e}xico}
\address{$^{*}$Facultat de F\'{\i}sica, Universitat de Barcelona. \\
Av. Diagonal 647, 08028, Barcelona, Spain}
\ead{isholek.fc@gmail.com, agustiperezmadrid@ub.edu
}
\begin{abstract}
We discuss the conditions for which the non-equilibrium work relation is valid by means
of thermodynamic and microscopic arguments.
\end{abstract}
\end{frontmatter}



\section{Introduction}

The question of dissipation and the approach to equilibrium in mesoscopic
and macroscopic systems has been a long standing problem since the early
years of statistical mechanics as a science. Already at the beginning of the
XX century Einstein and Ehrenfest were unsatisfied with the fundamental
principles of statistical mechanics postulated by Bolztmann and Gibbs, since
in their opinion those principles lack of a sound microdynamical basis.\cite%
{PAIS,COHEN1,COHEN2} Their main idea was that the real problem concerns the
approach towards the equilibrium state, and not equilibrium in itself. So,
the knowledge of the dynamics of the irreversibility and the approach to
equilibrium that follows enables one to draw the framework embodying
equilibrium as a limiting behavior.

This subject has acquired a renewed interest, mainly due to the
technological advances in biophysics and microrheology that make a more
detailed analysis of the question possible.\cite{liphart,mackintosh}

Related to this problem, attempts have been made to go beyond equilibrium in
order to establish some rules for extracting information from irreversible
processes. \cite{SENGERS,Groot,REVIEWjpcb,pnas, COHEN-GALLAVOTI,Jarzynski}
Recently it has been claimed that there are only a few known relations in
statistical mechanics that are valid for systems arbitrarily driven far from
equilibrium. One of these relations is the one proposed by Jarzynski \cite%
{Crooks,Jarzynski} for which equilibrium properties can be obtained from
non-equilibrium work measurements. For Hamiltonian isolated systems and
systems in contact with a heat bath through weak interactions, theoretical
justification for this assertion has been given in terms of ensembles of
trajectories represented by means of a phase space density.\cite{Jarzynski}
However, from the thermodynamic point of view this assertion seems to be
true only for very specific situations.\cite{COHEN3} As a consequence, it is
important to stipulate the conditions in which one can make use of this
relation. On the other hand, the knowledge of the range of applicability of
this and other non-equilibrium relations is deeply rooted in the
understanding of the microscopic bases of irreversibility.\cite%
{SENGERS,Groot,REVIEWjpcb,pnas,COHEN-GALLAVOTI}

In this article, we will use thermodynamic and statistical arguments to show that the
non-equilibrium work  relation is valid only in isothermal and near-equilibrium
conditions. Moreover, we will revise the validity of the original derivation
of this non-equilibrium work relation \cite{Jarzynski} and then establish a quantitative
criterion for determining these conditions by making use of a microscopic
theory involving the BBGKY (Bogoliubov-Born-Green-Kirkwood-Yvon) hierarchy of equations \cite{balescu}.

The paper is organized as follows. In Section \textbf{2}, we analyze the
non-equilibrium work  relation by considering the thermodynamic definition of the
minimum work done on a system to change its state. After this, in Sec. 
\textbf{3} we analyze the problem of irreversibility in the context of a
microscopic theory and propose the entropic time as a quantitative criterion
to distinguish quasistatic and nonquasistatic processes. Finally, the last
section is devoted to summarize and discuss our main results.

\section{Thermodynamics and cumulants}

In the  very general case, a system driven out of equilibrium does not satisfy
the condition of thermal equilibrium with the bath, since the temperature $T$
of the system differs from that of the bath $T_{0}$ \cite{ZWANZIG-LIBRO}. Among others, typical
systems illustrating this situation are glasses and supercooled colloidal
fluids.\cite{stillinger,weeks}

Since irreversible processes are present in this general case, the total work $R$
performed on the system in order to change its state satisfies the relation \cite{LANDAU} 
\begin{equation}
R\geq \Delta E-T_{0}\Delta S,  \label{W}
\end{equation}%
where $\Delta E$ and $\Delta S$ are the energy and the entropy variation of
the system in the process. Hence, in the case of a reversible process, from
Eq. (\ref{W}) one can infer that the minimum amount of work $R_{min}$
necessary to reverse the state of the system is given by 
\begin{equation}
R_{min}=\Delta E-T_{0}\Delta S.  \label{Wmin}
\end{equation}%
In the isothermal case, when the system is in contact with a heat bath, $T=T_{0}$, Eq. (\ref{Wmin}) can be used to
define the variation of the Helmholtz free energy of the system $\Delta F$ 
\begin{equation}
R_{min}=\Delta (E-T_{0}S)=\Delta F.  \label{Wmin-F}
\end{equation}%
Therefore, thermodynamics requires that the change of the free
energy of the system be equal to the work performed on it only in a
reversible process in which at least the initial and final states satisfy
the thermal equilibrium condition $T=T_{0}$.

In Ref. \cite{Jarzynski},
the following equation has been established 
\begin{equation}
\Delta F=-k_{B}T_{0}\ln \langle e^{-R/k_{B}T_{0}}\rangle ,
\label{Wmin-F-Jarynzki}
\end{equation}%
where $k_{B}$ is the Boltzmann constant and $\langle \,\rangle $ represents
the average over an ensemble of measurements of the work $R$.

In view of the thermodynamic relations (\ref{W})-(\ref{Wmin-F}), one may
legitimately ask about the conditions of validity of Eq. (\ref%
{Wmin-F-Jarynzki}). To answer this question, it is very important to
take into account that, in general, an arbitrary external perturbation will
produce an internal irreversible process in which the system approaches
equilibrium, in accordance with the Le Chatelier principle.\cite{LANDAU} 

For simplicity we will assume that work is done in isothermal conditions;
therefore, we have 
\begin{equation}
R=\Delta F+R_{diss},  \label{R}
\end{equation}%
where $R_{diss}$ is the dissipated work.\cite{LANDAU} Now, by taking the
exponential of Eq. (\ref{R}) one simply obtains 
\begin{equation}
e^{R/k_{B}T_{0}}=e^{\Delta F/k_{B}T_{0}}e^{R_{diss}/k_{B}T_{0}}.
\label{peaked}
\end{equation}%
After averaging this expression over the ensemble of measurements of $R$ one
obtains 
\begin{equation}
\langle e^{-R/k_{B}T_{0}}\rangle =e^{-\Delta F/k_{B}T_{0}}\langle
e^{-R_{diss}/k_{B}T_{0}}\rangle ,  \label{exp-gauss-av}
\end{equation}%
where we have taken into account that $\Delta F$ is a constant quantity between two equilibrium states.
Therefore, from Eq. (\ref{exp-gauss-av}) it follows that the relation $%
\langle e^{-R/k_{B}T_{0}}\rangle \simeq e^{-\Delta F/k_{B}T_{0}}$ is valid
only if the time scales characterizing the variation of the external
parameters are larger than the time scales characterizing the decay of the
irreversible processes taking place within the system, since then $%
R_{diss}\ll \Delta F$. A quantitative criterion for this time scale
will be given in the following section, where these irreversible processes
will be analyzed on microscopic basis.

For situations not far from equilibrium the implications of Eq. (\ref{peaked}) can be viewed in the context
of the linear response theory. Thus, by expanding the exponential containing $%
R_{diss}$ and averaging over the ensemble of measurements of $R$. Up to
first order,\cite{ZWANZIG-LIBRO} the result is 
\begin{equation}
\left\langle e^{R/k_{B}T_{0}}\right\rangle \simeq e^{\Delta F/k_{B}T_{0}} 
\left[ 1+\frac{\left\langle R_{diss}\right\rangle }{k_{B}T_{0}}+O(2)\right] ,
\label{linear-resp}
\end{equation}%
which seems to imply that Eq. (\ref{Wmin-F-Jarynzki}) does not contain the
first-order correction related to the response function of the system,
included here in $R_{diss}$. 

This conclusion is in accordance with the requirement that Eq. (\ref%
{Wmin-F-Jarynzki}) be valid for fluctuations of the work smaller than $%
k_{B}T $ as assumed in Ref.\cite{Jarzynski}: "the fluctuations in the work R must not be much greater than $%
k_{B}T $, if we are to have any hope of verifying Eq. (\ref{Wmin-F-Jarynzki}) experimentally...". A similar result can be obtained, for
example, by expanding the average $\langle e^{R/k_{B}T_{0}}\rangle $ around
the mean value $\langle R\rangle $ assuming that the distribution of work is
a Gaussian. Due to the symmetry of this distribution one obtains 
\begin{equation}
\langle e^{R/k_{B}T_{0}}\rangle \simeq e^{\langle R\,\rangle /k_{B}T_{0}} 
\left[ 1+\frac{ \langle R^{2}\rangle -\langle R\rangle ^{2} }{%
2(k_{B}T_{0})^{2}}+..\right] , \label{exp-gauss}
\end{equation}%
which as compared with Eq. (\ref{linear-resp}), leads to the estimate 
\begin{equation}
\frac{\left\langle R_{diss}\right\rangle }{k_{B}T_{0}}\sim \frac{%
\left\langle R^{2}\right\rangle -\langle R\rangle ^{2}}{2(k_{B}T_{0})^{2}}%
. \label{FDT}
\end{equation}%
This is precisely the result obtained in Ref. \cite{Jarzynski} in the paragraph that follows Eq. (12) in this reference. In addition, given that $\langle R\rangle =\Delta F+\langle R_{diss}\rangle $%
, by subtracting (\ref{linear-resp}) from (\ref{exp-gauss}) and taking
into account (\ref{FDT}),  one obtains $e^{\langle R_{diss}\,\rangle 
/k_{B}T_{0}}=1 $, which means that $\left\langle R_{diss}\right\rangle
\,/k_{B}T_{0}\simeq 0 $. Our linear response analysis, although it may seem naive, runs parallel to the one underlying in  the assumptions concerning the work fluctuations in Ref. \cite{Jarzynski}. A more elaborated linear response analysis has been done in Ref. \cite{ronis}, in which  the authors conclude that what \emph{the non-equilibrium work relation really  offers is not the thermodynamic free energy difference but merely an upper bound}: "most significantly, $\Delta A_j$  is not the free energy change predicted by thermodynamics.....". This assertion is relevant since it also indicates that Eq. (\ref{Wmin-F-Jarynzki}) gives only an approximate value of $\Delta F$, thus implying that this is not valid in general, contrary to what is commonly claimed. This is a widely extended misinterpretation which we attempt to clarify here.

The previous results highlight the thermodynamic implications of the
two assumptions underlying Eq. (\ref{Wmin-F-Jarynzki}). However, a deeper
analysis can be performed by noticing that $\langle e^{R/k_{B}T_{0}}\rangle $
is the moment-generating function of the probability distribution of $R$. 
\cite{cramer,kampen} Hence, in a general case we get more clarity if we
perform a series expansion of the quantity $-k_{B}T_{0}\ln \langle
e^{-R/k_{B}T_{0}}\rangle $,%
\begin{equation}
-k_{B}T_{0}\ln \langle e^{-R/k_{B}T_{0}}\rangle
=-k_{B}T_{0}\sum_{n=1}^{\infty }\frac{1}{n!}\frac{\kappa _{n}}{\left( -{%
k_{B}T_{0}}\right) ^{n}}.  \label{expansion}
\end{equation}%
where $\kappa _{n}$\ is the n-th order cumulant. 

Using now the theory of Thiele semi-invariants, \cite{cramer,zwanzig}
we can define the $n$-th order cumulant in terms of the moments of ${%
\left\langle R^{k}\right\rangle }$. For the first few values of $n$, we have 
\begin{eqnarray}
\mathbf{\ }\kappa _{1} &=&{\left\langle R\right\rangle }   \\
\kappa _{2} &=&{\left\langle R^{2}\right\rangle -\langle R\rangle ^{2}} 
 \\
\kappa _{3} &=&{\left\langle R^{3}\right\rangle -3\langle R}^{2}{\rangle
\left\langle R\right\rangle +2\left\langle R\right\rangle }
\label{cumulants} \\
\kappa _{4} &=&{\left\langle R^{4}\right\rangle -4\left\langle
R^{3}\right\rangle \left\langle R\right\rangle -3\left\langle
R^{2}\right\rangle +12\langle R}^{2}{\rangle \left\langle R\right\rangle
-6\left\langle R\right\rangle }.  
\end{eqnarray}%
It can be shown\cite{kampen} that for a Gaussian distribution, $%
\kappa _{n}=0$ for $n>2$. So, the value of the
quantity $-k_{B}T_{0}\ln \langle e^{-R/k_{B}T_{0}}\rangle $ 
depends on the probability distribution, and therefore cannot coincide with the thermodynamic free energy since the thermodynamic free energy must be independent from the statistics. As we have shown above, this coincidence occurs only in the Gaussian case when the system is not arbitrarily far away from equilibrium (\textit{i.e.}, in the fluctuation-dissipation regime). A cumulant analysis has been also performed in several papers, in Ref. \cite{Jarzynski} itself or in Refs.  \cite{ronis, presse} just to cite a few exemples. Nonetheless, the approach in these works differs from ours since in these the non-equilibrium work relation is taken for granted and the cumulant expansion serves to test how great the statistic must be to compute the free energy. For us however, the cumulant expansion is less restrictive since the only thing that this gives us for sure is information about the probability distribution of work fluctuations and very little else.

Let us illustrate our point with a simple example which shows that Eq. (\ref{Wmin-F-Jarynzki}) is not general. A typical distribution in non-equilibrium situations is the
chi-squared or gamma distribution, \cite{papoulis} for which%
\begin{eqnarray}
-k_{B}T_{0}\ln \langle e^{-R/k_{B}T_{0}}\rangle =-k_{B}T_{0}\ln \langle
e^{-\left( R/\Delta F\right) \left( \Delta F/k_{B}T_{0}\right) }\rangle = 
\\
\frac{k_{B}T_{0}}{2}\frac{{\left\langle R\right\rangle }}{\Delta F}%
\sum_{n=1}^{\infty }\frac{1}{n}\left( 2\frac{\Delta F}{{k_{B}T_{0}}}\right)
^{n},  \label{expansion_2}
\end{eqnarray}%
where work has been measured in units of $\Delta F$. The sum in Eq.
(\ref{expansion_2}) is a Mercator series which is
convergent when $2\Delta F/{k_{B}T_{0}<1}$. Hence in this case, we obtain  
\begin{eqnarray}
-k_{B}T_{0}\ln \langle e^{-R/k_{B}T_{0}}\rangle  &=&-\frac{k_{B}T_{0}}{2}%
\frac{{\left\langle R\right\rangle }}{\Delta F}\ln \left( 1-2\frac{\Delta F}{%
{k_{B}T_{0}}}\right) \simeq  \\
{\left\langle R\right\rangle }=  
{\left\langle R_{diss}\right\rangle }+\Delta F  &=&\Delta
F\left( 1+\frac{{\left\langle R_{diss}\right\rangle }}{\Delta F}\right)\neq \Delta F.
\label{sum}
\end{eqnarray}%
Therefore, clearly the non-equilibrium work relation is not satisfied in this case. Thus, by considering nonequilibrum situations where the
distribution of $R_{diss}$ can be fitted to a chi-square distribution 
\begin{equation}
F(x)dx={1}/{[2\Gamma (f/2)]}\left( {x}/{2}\right) ^{f/2-1}e^{-x/2}dx,
\label{chi-square}
\end{equation}%
with variance $\sigma _{x}^{2}=2f$ and mean $m_{x}=f$, with $x\equiv
R_{diss}/k_{B}T_{0}$ and $f\sim \left\langle x\right\rangle $, 
we can compute%
\begin{equation}
\left\langle e^{-x}\right\rangle =\int_{0}^{\infty }e^{-x}F(x)dx=\left( {1}/{%
3}\right) ^{f/2}\neq 1  \label{average}
\end{equation}%
Hencen, since according to Eq. (\ref{average}) $\left\langle e^{-x}\right\rangle\neq 1$ one concludes that $\langle e^{-R/k_{B}T_{0}}\rangle\neq e^{-\Delta F/k_{B}T_{0}}$ which violates the conditions we have found through Eqs. (\ref{linear-resp})-(\ref{FDT}). These results clearly show that unlike what is claimed in the current literature \cite{Crooks}, for arbitrary large fluctuations the non-equilibrium work relation is not applicable and thus is only valid for equilibrium fluctuations. 

Experimentally, the mean value $\langle \,R\rangle $ is obtained through an
arithmetic mean. A Gaussian distribution centered in $\left\langle
R\right\rangle $ is used to account for data dispersion. Since the number of
experiments performed is finite, this distribution corresponds to the
statistical distribution of frequencies of the measured values of $R$.
However, as is emphasized in Fig. \textbf{1}, this statistical distribution
does not necessarily coincide with the limiting distribution whose maximum
value corresponds to the free energy difference $\Delta F$. As a
consequence, it is quite possible that the measurements will have values
near those of the limiting distribution, and any difference will be
interpreted as a statistical error and not due to other kind of bias.

To conclude our analysis of Eq. (\ref{Wmin-F-Jarynzki}), we will consider a
system far from equilibrium which is in contact with a heat bath,
whose entirety constitutes an isolated system\cite{Jarzynski}. Under these
conditions, it is very important to note that the temperature $T\propto
\beta ^{-1}$ is a function of the energy, \cite{Huang, Pathria} regardless of the
size of the system \cite{hill-small}. There is also more recent literature emphasizing this fact, \cite{evans, powles, rugh}. Therefore, for a time-dependent
Hamiltonian system, $T$ is not a thermodynamic variable in itself but rather a
time-dependent parameter. Hence, the use of a constant equilibrium
temperature $T_{0}\propto \beta _{0}^{-1}$ as a prefactor in the
exponentials of Eqs. (7) and (8) of Ref. \cite{Jarzynski} is not appropriate, 
since its time dependence must be taken into account.  In our opinion this non-adequate treatment of the temperature is also present in Ref. \cite{Jarzynski_2}. It should be noted that our perception about the temperature is coincident 
with the same objection raised in Refs. \cite{COHEN3, cohen_4}.
In view of this, we may follow Ref. \cite{Jarzynski} in order to write
\begin{eqnarray}
f(z;t)e^{-\beta _{0}w(z;t)} &=&Z_{0}^{-1}e^{-\beta
_{0}H_{0}(z_{0})}e^{-\beta _{0}w(z;t)} \\
&=&Z_{0}^{-1}e^{-\beta _{0}H_{\lambda }(z;t)},  \label{particionJ}
\end{eqnarray}%
where $w(z;t)=H_{\lambda }(z;t)-H_{0}(z_{0})$ is the work done along a
trajectory $z(z_{0},t)$ with initial condition $z_{0}$. However, since for a
finite switching time $t_{s}$ $\beta (t_{s})=\beta _{1}\neq \beta _{0}$, one
obtains 
\begin{equation}
\int dzexp[-\beta _{0}H_{1}]\neq Z_{1}\equiv \int dzexp[-\beta _{1}H_{1}], \label{particionJ_2}
\end{equation}
in contradiction with the result given in Ref. \cite{Jarzynski}.  
Additionally, it must be emphasized that the definition of work $w$ in Eq.
(\ref{particionJ}) is not consistent with the one corresponding to
thermodynamic processes. The consistency between these two definitions
arises only when the processes considered are slow enough, \textit{i.e.},
for adiabatic processes for which the entropy is a constant and therefore
constitute reversible processes.\cite{LANDAU} 

As a consequence of the previous analysis, it follows that it is not
correct to identify the quantity $w$\cite{Jarzynski} with the thermodynamic work $R$ which,
through Eqs. (\ref{Wmin-F}) and (\ref{R}), constitutes the definition of the
free energy difference $\Delta F$. Thus, after using Eq. (\ref{exp-gauss-av}%
), one can write 
\begin{equation}
\Delta F=-k_{B}T_{0}\ln \left[ \frac{\langle e^{-R/k_{B}T_{0}}\rangle }{%
\langle e^{-R_{diss}/k_{B}T_{0}}\rangle }\right] ,  \label{jar-gen}
\end{equation}%
which is the general relation arising from thermodynamic arguments.

\section{Microscopic analysis: The entropic time}

The discussion in the previous section allows us to conclude that,
far away from equilibrium, the approximations made in order to derive
non-equilibrium work relation's (\ref{Wmin-F-Jarynzki}) are no longer valid. Hence, the
evaluation of the dissipated work becomes necessary.

In order to achieve this objective and to obtain a quantitative
criterion establishing when Eq. (\ref{Wmin-F-Jarynzki}) can be applied, we
will write the work dissipated by a system in contact with a heat bath and
due to the action of an external force $X_{j}$ in the form\cite{KATCHALSKI} 
\begin{equation}
R_{diss}=X_{j}\Delta \xi _{j},  \label{Rdis-kat}
\end{equation}%
where $\xi _{j}$ is an internal parameter characterizing the state of the
system. Since in view of$~$Eq. (\ref{Rdis-kat}) $\ R_{diss}$ is related to
the entropy production of the system $\dot{\sigma}$, we may write Eq. (\ref%
{exp-gauss-av}) 
\begin{equation}
\langle e^{-R/k_{B}T_{0}}\rangle \simeq e^{-\Delta F/k_{B}T_{0}}\left[ 1-%
\frac{\dot{\sigma} \tau_s}{k_{B}} +O(2)\right] ,  \label{linear-sigma}
\end{equation}%
where $\tau_s $ is an elapsed time. From this equation it follows that it is
necessary to estimate the characteristic relaxation time of the entropy
production in order to determine when Eq. (\ref{Wmin-F-Jarynzki}) is valid.

This task can be accomplished by means of a microscopic theory that we will
summarize here. For an isolated N-body system, in Ref. \cite{JSM-Agustin, JCP-Agustin, PHY-Agustin}
it was postulated the non-equilibrium entropy $S$ in the BBGKY scenario as a
functional of the set of $s$-particle reduced distribution functions ($s\leq
N$), represented in the distribution vector $\mathbf{f}$ 
\begin{eqnarray}
S &=&-k_{B}tr\left\{ \mathbf{{f}\ln \left( {f}_{eq}^{-1}{f}\right) }\right\}
+S_{eq}=\,\,\,\,\,\,\,\,\,\,\,\,  \label{gibbs-postulate} \\
&&-k_{B}\sum_{s=1}^{N}\frac{1}{s!}\int f_{s}\ln \frac{f_{s}}{f_{eq,s}}
dx_{1}...dx_{s}+S_{eq}, 
\end{eqnarray}%
which generalizes the Gibbs entropy postulate\cite{Groot,REVIEWjpcb,pnas}.
In this relation $S_{eq}$ is the (thermodynamic) entropy at equilibrium and $%
\mathbf{{f}_{eq}} $ is the distribution vector that corresponds to the
equilibrium state. Hence, $
\mathbf{{f}_{eq}} $ is the equilibrium solution of the BBGKY hierarchy, in other words this is a solution of the YBG (Yvon-Born-Green) hierarchy \cite{balescu, hansen}, and therefore  is not a Boltzmann-like function. Additionally, the non-equilibrium entropy (\ref%
{gibbs-postulate}) reaches its maximum value at equilibrium, when $\mathbf{{f%
}={f}_{eq}}$.

In the framework of the BBGKY description, the system is assumed to be a
mixture of $s$-particle interacting fluids in the phase space and each fluid
is made up of particle-clusters of equal size.\cite{JSM-Agustin} As a
consequence of the interaction among those fluids a compressible multiphase
flow is established in the phase space, so it has been proved \cite%
{JSM-Agustin} that Eq. (\ref{gibbs-postulate}) is a monotonically increasing
function in time that properly describes the regression to equilibrium of a
system originally under non-equilibrium conditions.

In Ref. \cite{JSM-Agustin} , 
it was also shown that the dynamics of the non-equilibrium entropy (\ref%
{gibbs-postulate}) follows from the dynamics of the distribution vector $%
\mathbf{f}$ given through 
\begin{equation}
\frac{\partial \mathbf{f}(t)}{\partial t}=\mathbf{\mathcal{L}{f}(t),}
\label{liouville}
\end{equation}%
which is the generalized Liouville equation expressing the BBGKY hierarchy
in a compact way, where $\mathcal{L}$ is the generalized Liouvillian. As it
is known, Eq. (\ref{liouville}) is obtained by projecting adequately the
Liouville equation onto each one of the $s$-particle phase space. Therefore,
using Eqs. (\ref{gibbs-postulate}) and (\ref{liouville}) one can compute the
entropy production 
\begin{equation}
\dot{\sigma}\equiv \frac{\partial S}{\partial t}=-\frac{1}{T}\sum_{s=1}^{N}%
\frac{1}{s!}\sum_{j=1}^{s}\int J_{j}\Delta {\mathcal{F}}_{j}dx_{1}...dx_{s},
\label{entrop-prod}
\end{equation}%
where $\Delta {\mathcal{F}}_{j}$ is the averaged non-equilibrium force
between the $j$-th particle and the particles of the remaining fluids.
Moreover, $J_{j}$ arises from deviations of the distribution function with
respect to equilibrium.

From Eq. (\ref{entrop-prod}) we may estimate the relaxation time of the
irreversible processes taking place in the system. To this end, note that
the product $J_{j}dx_{1}...dx_{s}$ has dimensions of velocity whereas $%
\Delta \mathcal{F}$ has dimensions of force. Introducing the characteristic
velocity $\overline{v}$ and taking into account that $\Delta \mathcal{F}$
contains the interaction forces among the components of the system, then $%
\Delta {\mathcal{F}}\sim \phi _{0}/r_{0}$ with $\phi _{0}$ and $r_{0}$ being
the characteristic energy and the characteristic length of the interaction
potential. Thus, the entropic time scale $\tau _{s}={m \overline{v}r_{0}}/{\phi _{0}}$ (with $m$ the mass of the system) related to the approach to equilibrium is the time associated to
the relaxation of the non-equilibrium entropy to its equilibrium value. If
we assume that $\overline{v}\sim \sqrt{k_{B}T/m}$ with $m$ the mass of the
system, we finally obtain 
\begin{equation}
\tau _{s}\simeq \sqrt{mk_{B}T}\,({r_{0}}/{\phi _{0}}),  \label{time}
\end{equation}%
which has the expected physical limits at low and high temperatures. Eq. (%
\ref{time}) implies that experimental measurements can be interpreted in the
context of Eq. (\ref{Wmin-F-Jarynzki}) only if the time elapsed before
taking a measurement in each point of the trajectory is larger than $\tau
_{s}$. It is important to note that the dependence on mass of Eq. (\ref{time}%
) implies that the time of decay of irreversible processes in nanoscale
systems could be very short, thus making them appropriate candidates to make
the misleading interpretation that Eq. (\ref{Wmin-F-Jarynzki}) is a far-from
equilibrium relation.

Here, it is convenient to consider that the thermodynamics of small systems,\cite{hill-small} introduces corrections to the thermodynamic variables and state functions which make them different from their corresponding macroscopic counterparts. The magnitude of these corrections depend directly on the size of the ensemble of systems which is considered. 
\cite{hill-small} From a dynamical point of view, for small systems these deviations from the macroscopic behavior are not small and may affect, in general, not only the extensive quantities but also the intensive ones (as, for example, the Massieu function). This fact is important, since these fluctuations modify the dynamics in such a way that may introduce multiplicative stochastic noises in the corresponding evolution equations for the variables determining the state of the system. Often, these multiplicative fluctuations lead to polynomial corrections to the local Gaussian distributions associated with the fluctuating quantities. As shown in the previous section, these corrections will affect the cumulant expansion of the corresponding distribution. The assumption that for arbitrary nonequilibrium processes the higher order terms of Eqs. (8)-(13) cancel between them in general, seems to be incorrect.

In the previous paragraph we analyzed the irreversible behavior of an
isolated system. However, since most of the experiments are performed in the
presence of a bath, we should generalize our analysis.

In the case when the system is in contact with a bath, it is necessary to
extend the full phase-space in order to account for the presence of the
bath. In this framework, the volume elements of the new phase space are
conserved and therefore the Liouville theorem is satisfied. Thus, we will
include a set of pairs of conjugated generalized coordinates $(\lambda ,\dot{%
\lambda})$ so that the Hamiltonian is now a function of $\lambda $ and $\dot{%
\lambda}$, $H(p,q,\lambda ,\dot{\lambda})$.\cite{LANDAU} Here, $\lambda $
represents the set of external parameters determining the state of the
system and accounting for the interaction with the bath and $\dot{\lambda}$
the set of conjugated generalized momenta (which can be also understood as
external currents).

Therefore, in this case equation (\ref{liouville}) is rewritten in the form 
\begin{equation}
\frac{d_{\lambda}\mathbf{f}(t)}{dt}\equiv \left( \frac{\partial }{\partial t}%
+\dot{\lambda}\frac{\partial }{\partial \lambda }\right) \mathbf{{f}(t)=%
\mathcal{L}{f}(t),}  \label{liouville-lambda}
\end{equation}%
where $d_{\lambda }/dt$ is the convective derivative associated with the
external currents. Following the procedure that bring us Eq. (\ref%
{entrop-prod}), equations (\ref{gibbs-postulate}) and (\ref{liouville-lambda}%
) lead to 
\begin{eqnarray}
\frac{d_{\lambda }S}{dt}= %
\dot{\sigma}_{\lambda } +\frac{\dot{Q}}{T_{0}},
\label{entrop-prod-lambda}
\end{eqnarray}%
with $\ \dot{Q}$ being the rate of heat exchange between the system and the
bath. In addition, $\dot{\sigma}_{\lambda }$ stands for the entropy
production of the system at a given value of the external parameter $\lambda 
$ imposed by the bath. In this case Eq. (\ref{linear-sigma}) is written as%
\begin{equation}
\langle e^{-R/k_{B}T_{0}}\rangle\simeq e^{-\Delta F/k_{B}T_{0}}\left[ 1-%
\frac{\dot{\sigma}_{\lambda }\tau_{s}}{k_{B}} +O(2)\right] .
\label{linear-sigma_2}
\end{equation}%
In addition, when the internal irreversible processes in the system have
decayed ($t \geq \tau _{s}$), Eq. (\ref{entrop-prod-lambda}) reduces to the
well known Clausius relation $\delta S=\delta Q/T_{0}$, valid for
quasiestatic processes.

\section{Conclusions}

In this article, we have analyzed the non-equilibrium work  relation on the grounds of
equilibrium thermodynamics and in terms of the underlying microscopic
dynamics based on the BBGKY description. Our analysis is applicable no
matter how distant from equilibrium the system is.

We have shown that, in general, in order to be valid non-equilibrium work relation's (\ref{Wmin-F-Jarynzki})
should include a correction taking into account  the internal dissipation in the
system. Additionally, we have shown that Eq. (\ref{Wmin-F-Jarynzki})
corresponds to the zero order term in a linear response analysis and thus 
is merely a near equilibrium result (\textit{i.e.}, it applies when the fluctuations are gaussian, in the fluctuation-dissipation regime).

From the microscopic dynamics of the system, given through the generalized
Liouville equation and from the entropy postulate (\ref{gibbs-postulate}),
we have computed the entropy production and provide an estimate of the
characteristic time scale for the irreversible process taking place in the
system, \emph{the entropic time} $\tau _{s}$. This quantity depends on the
mass, the temperature, the characteristic length and the magnitude of the
interaction potential and therefore establishes the criterion that must be
satisfied by a process in order to be quasiestatic. This phase-space
analysis is the most appropriate for small systems characterized by smooth
phase-space distribution functions.

Once again to avoid misunderstandings we emphasize  that our microscopic analysis is based on a generalization of the Gibbs entropy postulate defined through Eq. (\ref{gibbs-postulate}) which is a functional of the distribution vector $\mathbf{f}(t)$. This distribution vector represents the whole set of s-particle reduced distribution functions with $s=0,.....,N$ whose dynamics is given by the generalized Liouville equation (\ref{liouville}). Therefore, our entropy  (\ref{gibbs-postulate}) is not a constant of motion under the generalized Liouville dynamics  (\ref{liouville}) as repeatedly has been shown in several applications \cite{JSM-Agustin, JCP-Agustin, PHY-Agustin, PHY_2-Agustin}.
Also incorporated in the theory is the presence of a bath which introduces a
drive. This has been carried out by including the degrees of freedom of the
bath into the full phase space through the external parameters used to
characterize the state of the system. In this way, the volumes of the phase
space are conserved and the generalized Liouville equation (\ref%
{liouville-lambda}) remains valid. These external parameters originate
currents entering the generalized Liouville equation and consequently also
in the distribution function. In this scenario, the non-equilibrium entropy
depends on the \emph{external} currents through the distribution functions.

The present discussion establishes $\tau _{s}$ as a useful quantitative
criterion to implement experiments in good agreement with the theory.

\textbf{ACKNOWLEDGMENTS} 

We acknowledge Prof. R. F. Rodr\'iguez for interesting discussions. This
work was supported by UNAM-DGAPA under the grant IN-108006.

\clearpage
Fig. 1. Limiting (solid line) and frequency (dashed line) distributions as a
functions of the reduced work $R^{\ast }=R/k_{B}T$. The figure shows that $%
\langle R\rangle \simeq \Delta F$ only when the number of experimental
measurements $N\rightarrow \infty $ and the internal irreversible processes
of the system have relaxed. \clearpage
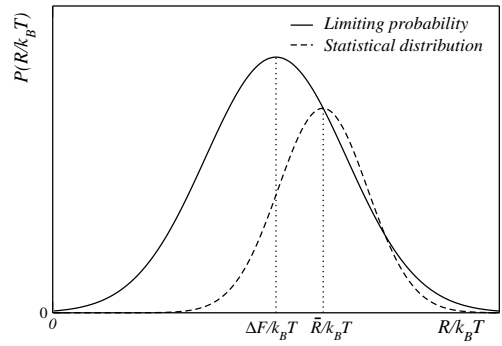
\begin{figure}[tbp]
{}
\par
\centering
\mbox{\resizebox*{6.5cm}{!}{\includegraphics{Fig1-AIB.eps}} }
\par
{\footnotesize {\ } \vspace{0.0cm} }
\caption{\,}
\label{velpromedio}
\end{figure}

\end{document}